\documentclass[prl,twocolumn,showpacs,superscriptaddress,amsmath]{revtex4-1}
\usepackage{amssymb}
\usepackage{graphicx}
\usepackage{dcolumn}
\usepackage{bm}
\usepackage[utf8]{inputenc}
\usepackage{mathtools}
\usepackage[version=3]{mhchem}
\usepackage{verbatim}
\usepackage{units}
\usepackage[dvipdfm, pdfstartview=FitH, CJKbookmarks=true, bookmarksnumbered=true,
bookmarksopen=true, colorlinks, pdfborder=001, linkcolor=blue, anchorcolor=blue, citecolor=blue]{hyperref}
\usepackage{color}

\begin{document}


\title{Observation of Landau quantization and standing waves in HfSiS}

\author{L. Jiao}
\affiliation{Max-Planck-Institute for Chemical Physics of Solids, N\"othnitzer Str. 40, 01187 Dresden, Germany}
\author{Q. N. Xu}
\affiliation{Max-Planck-Institute for Chemical Physics of Solids, N\"othnitzer Str. 40, 01187 Dresden, Germany}
\author{Y. P. Qi}
\affiliation{Max-Planck-Institute for Chemical Physics of Solids, N\"othnitzer Str. 40, 01187 Dresden, Germany}
\affiliation{School of Physical Science and Technology, ShanghaiTech University, Shanghai 201210, China}
\author{S.-C. Wu}
\affiliation{Max-Planck-Institute for Chemical Physics of Solids, N\"othnitzer Str. 40, 01187 Dresden, Germany}
\author{Y. Sun}
\affiliation{Max-Planck-Institute for Chemical Physics of Solids, N\"othnitzer Str. 40, 01187 Dresden, Germany}
\author{C. Felser}
\affiliation{Max-Planck-Institute for Chemical Physics of Solids, N\"othnitzer Str. 40, 01187 Dresden, Germany}
\author{S. Wirth}
\email{Steffen.Wirth@cpfs.mpg.de}
\affiliation{Max-Planck-Institute for Chemical Physics of Solids, N\"othnitzer Str. 40, 01187 Dresden, Germany}

\date{\today}

\begin{abstract}
Recently, HfSiS was found to be a new type of Dirac
semimetal with a line of Dirac nodes in the band structure. Meanwhile, Rashba-split surface states are
also pronounced in this compound. Here we report a systematic study of HfSiS by scanning tunneling
microscopy/spectroscopy at low temperature and high magnetic field.
The Rashba-split surface states are characterized by measuring Landau quantization and standing waves, which
reveal a quasi-linear dispersive band structure. First-principles calculations based on density-functional 
theory are conducted and compared with the experimental results. Based on these investigations,
the properties of the Rashba-split surface
states and their interplay with defects and collective modes are discussed.
\end{abstract}
\maketitle

Over the past few decades, spin-orbit coupling (SOC) played a role in understanding the
intricate properties of electrons in a solid. Although SOC, normally, only adds a small perturbation to
the Hamiltonian of a system, in some materials it can be significant. One well-known example
is the Rashba SOC, which results from
inversion symmetry breaking~\cite{Rashba}. At the sample's surface or interface, Rashba SOC can be markedly
enhanced due to the large electric-field gradient ~\cite{Soumyanarayanan}. Consequently, the spin degeneracy
will be lifted by the strong SOC, which induces many novel phenomena, such as Rashba-split surface states and the
geometrical (Berry) phase~\cite{Soumyanarayanan,Manchon}. Recently, the topological nature of solids provided another hot topic
related to the SOC. In a topological insulator, strong SOC can invert the bulk valence and conduction bands at the surface,
which is a prerequisite for the formation of topological surface states~\cite{HasanRMP,QiRMP}. Soon after
the discovery of topological insulators, Dirac semimetals with various exotic phenomena and promising
applications were heavily investigated, including three-dimensional (3D) Dirac semimetals~\cite{Liu2014} and Weyl semimetals~\cite{Xu2015}.

More recently, crystals belonging to non-symmorphic space groups have attracted new research interest. The non-symmorphic symmetry
can induce a new symmetry-protected topological phase of matter, the so-called Dirac nodal line ~\cite{Kim,Yu,BianPRB,Fang}. To date, several materials have been investigated,
such as CaAgAs~\cite{Yamakage,Takane1}, PbTaSe$_2$, PbTaSe$_2$~\cite{Bian}, PtSn$_4$~\cite{Wu},
$M$Si$X$ ($M$ = Zr/Hf; $X$ = S/Se/Te)~\cite{Schoop,Neupane,Takane,Chen,HuPRL,Topp,Hosen,Pezzini,Bu}, Ca$_3$P$_2$~\cite{Chan}, InBi~\cite{Ekahana}, CaTe~\cite{Du}, and monolayer Cu$_2$Si~\cite{Feng}.
In these Dirac nodal-line semimetals, the linear bands do not cross
at discrete point(s) but form a continuous line(s) in the momentum space. Among these materials, the $M$Si$X$ family
provides a good opportunity to study the interplay between SOC and non-symmorphic symmetry due to their high tunability,
chemical stability, and quasi-2D crystal structure in favor of surface-sensitive techniques. Band-structure calculations and
angle-resolved photoelectron spectroscopy (ARPES) measurements have mapped out their band structure~\cite{Xu,Schoop,Neupane,Takane,Chen}.
An extraordinary property of this family is that the linear band dispersion region can be as large as 2 eV.
Additionally, unusual magnetotransport properties were experimentally observed~\cite{HuPRL,Wang,Singha,Ali,Kumar,Matusiak}.
On the other hand, the Rashba parameter, which is used to estimate the
strength of Rashba splitting, has been reported to be as large as $\sim$3 eV {\AA}~\cite{Takane,Chen}.
This value is nearly one order larger than in normal metals, like Au
and Sb~\cite{Soumyanarayanan2015}.
However, systematic investigations of the surface
states and their interplay with impurities and magnetic field using surface-sensitive methods are still lacking.
Especially, theoretical investigations proposed
that the quantization of the Dirac Landau levels are flat and should be
non-dispersive, which is of particular interest~\cite{Rhim}.

Here, we investigate HfSiS by utilizing scanning tunneling microscopy/spectroscopy (STM/STS)
down to a base temperature of 0.35~K and
up to a magnetic field of 12~T. A small bias voltage is applied to the sample with respect to
the tungsten tip. Clear Landau quantization of the
local density of states (LDOS) is observed in a high magnetic field. On the other hand, the dispersion
of the Rashba-split surface states is visualized by measuring the standing waves along a step-edge.
These results are consistent with our band-structure calculations.

\begin{figure}[tb]\centering
\includegraphics[width=7cm]{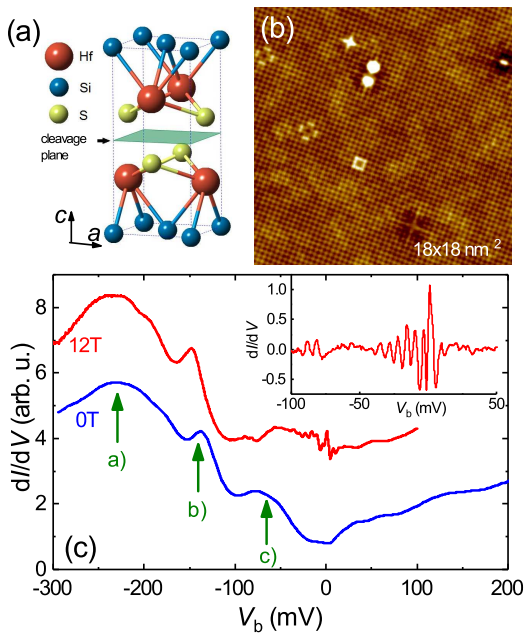}
\caption{(a) Crystal structure of HfSiS. The green
plane indicates the cleavage position. (b) 18 $\times$ 18~nm$^2$ STM topography of the cleaved
S-terminated surface of HfSiS. The image was taken at a tunneling condition of bias voltage $V_b$ = 300~mV,
current set point $I_{set}$ = 0.8~pA, and sample temperature $T$ = 0.35~K. (c) Typical STS spectra measured on a
clean site in a magnetic field of 0 and 12~T with $B \parallel c$. The spectrum measured at 12~T is
offset by 2 arb. u. Tunneling conditions used for all the STS measurements are $V_b$ = 0.1~V and
$I_{set}$ = 0.5~nA, while the modulation voltage $V_{mod}$ = 3~mV. The inset shows the spectrum obtained
at 12~T after subtracting a background signal.}
\label{Fig1}
\end{figure}

Single crystals of HfSiS are grown by the chemical vapor transport method in a standard two-step process
described elsewhere~\cite{Kumar}. The quality of the sample was carefully checked by x-ray diffraction
(XRD) and energy-dispersive x-ray (EDX) spectroscopy. Our XRD measurements show that HfSiS
crystallizes in the PbFCl-type structure of $P$4/$nmm$ space group (No.~129), as shown in
Fig.~\ref{Fig1} (a).
The lattice parameters are $a$~=~$b$~=~3.52~{\AA} , $c$~=~8.00~{\AA}. In this
compound, Hf-S double layers are sandwiched between Si square nets, forming the nonsymmorphic symmetry.
On the other hand, the relatively weak interactions between S layers provide a natural cleavage
plane. By cleaving the sample at low temperature (below 20~K) and in ultra high vacuum ($\sim$10$^{-9}$~Pa), a shiny
S-terminated surface is exposed. Figure~\ref{Fig1}(b) displays a typical surface topography of
HfSiS obtained by STM. The inter-atomic S-S distance in the STM topography was found to be 3.50(7)~{\AA},
which is consistent with the lattice parameter obtained from the XRD measurements.
In addition to the well-ordered lattice of sulfur atoms, there are several types of defects
that originate from impurities or vacancies on the top or the second layer. Comparing
the number of defects to the number of unit cells within the field of view, the density of defects is rather small,
confirming high sample quality.
In the following, we show that the perturbation to the surface states induced by
the impurities can extend up to an order of about ten unit cells,
indicating that these defects may have a significant influence on the surface property.

The intrinsic electronic properties of HfSiS are studied by STS measurement on a
clean site that is away from any defects. Figure~\ref{Fig1}(c) shows the differential conductance
spectra acquired by a standard lock-in technique at 0.35~K. The blue and red curves are obtained
at a magnetic field of 0 and 12~T, respectively. At zero field, the broad ``V"-shaped d$I$/d$V$-curve around Fermi level ($E_F$)
manifests the semi-metallic nature of HfSiS. However, there are pronounced peaks adding additional features to
the curve at negative bias voltage ($V_b$). The very broad hump a) located around $-$230~mV is also
manifested in the d$I$/d$V$-curve of ZrSiS~\cite{SankarSR}, which is understood in terms of van Hove singularities.
\begin{figure}[tb]\centering
\includegraphics[width=7cm]{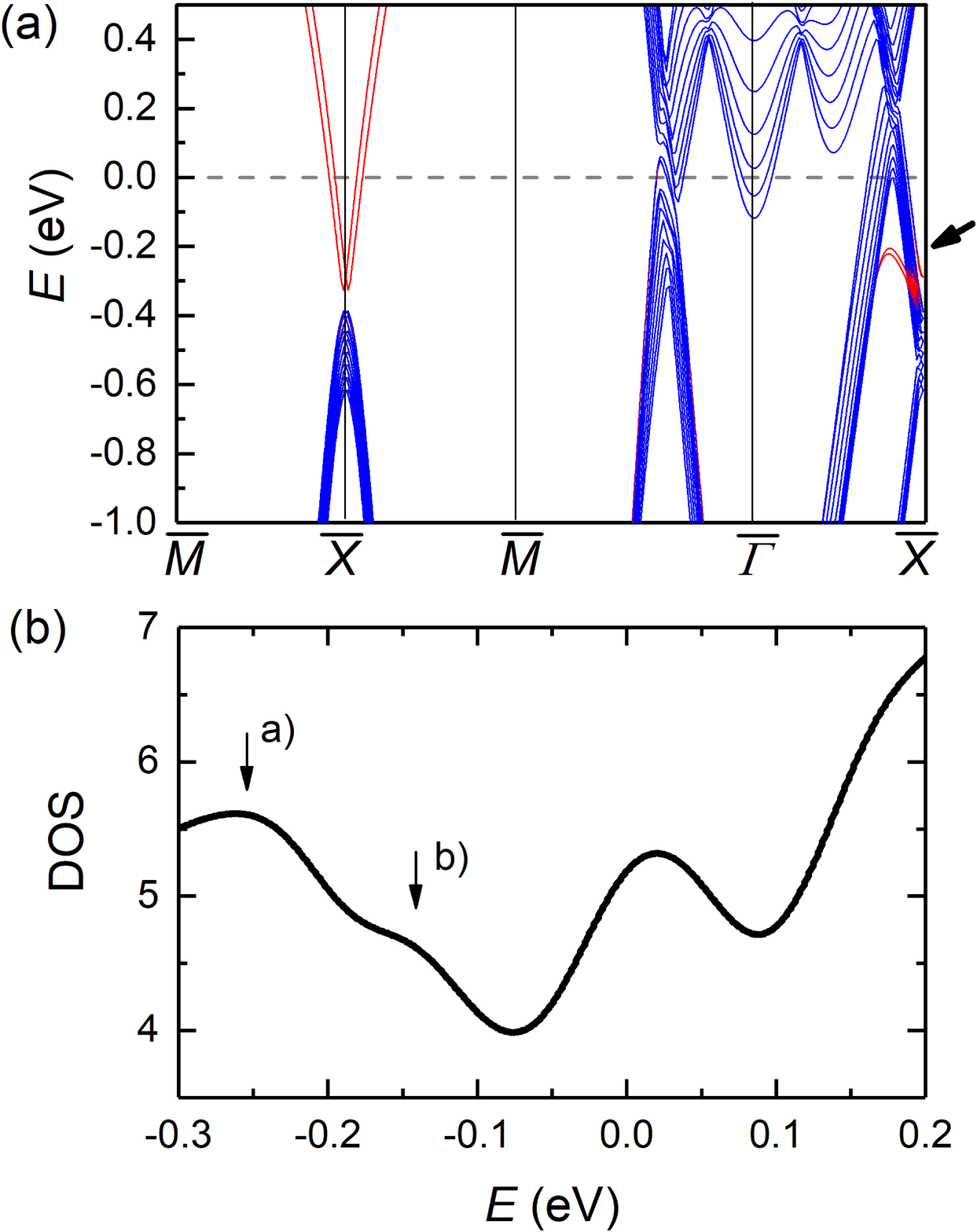}
\caption{(a) Slab calculation of the band structure of HfSiS along certain high symmetry directions.
The red and blue bands are surface and bulk states, respectively. Black arrow indicates the van Hove
singularities corresponding to peak a) in the bottom panel.(b) Calculated DOS
of the slab structure of HfSiS from $-$0.3 to 0.2 eV.}
\label{Figcal}
\end{figure}

To further analyze our STS result, we also conducted a first-principles calculation based
on the density functional theory (DFT). Here, we applied the Vienna ab-initio simulation package (VASP)
and used the generalized gradient approximation (GGA) of Perdew-Burke-Ernzerhof (PBE) as the
exchange-correlation potential. A ten-cell-thick slab structure of HfSiS along the $z$-direction
with 10 {\AA} vacuum was used in order to capture both bulk and surface properties. The calculated
band dispersion as well as the density of states (DOS) of this slab structure are presented in
Fig.~\ref{Figcal}. The Rashba-split surface states show a quasi-linear dispersion in Fig.~\ref{Figcal}(a) (red curves), which
yields a Fermi velocity of 4.39 eV {\AA} (6.67 $\times$ 10$^5$ m/s).
\begin{figure*}[tb]\centering
\includegraphics[width=11cm]{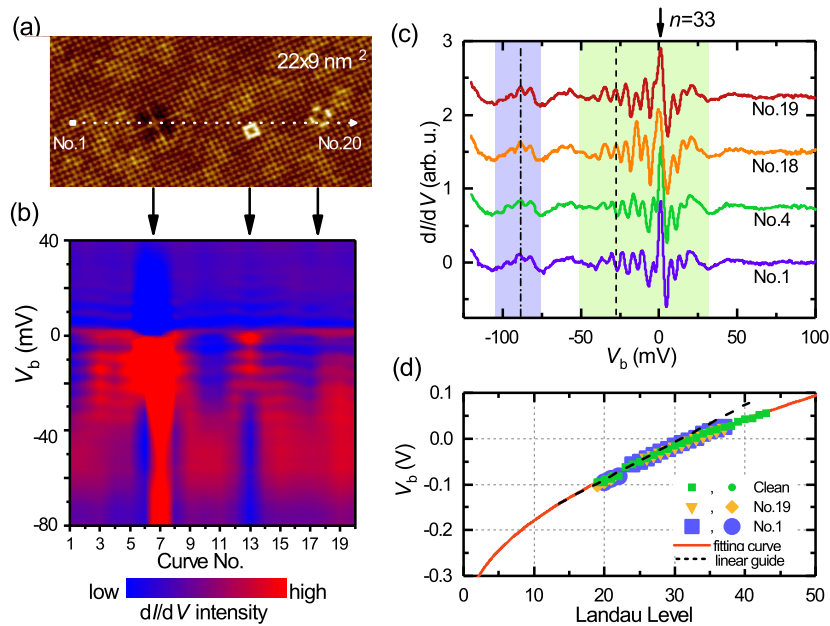}
\caption{(a) Topography of the surface of HfSiS with three
different types of defects (marked as $\bigoplus$, $\square$, and $\maltese$ in the main text).
(b) Contour plot of the 20 d$I$/d$V$-curves measured along the 20 nm white dot line in (a).
The spectra were obtained at 0.35~K and 12~T. (c) Selected
d$I$/d$V$-curves with a background signal subtracted. Curves are equally offset for clarity.
(d) The Landau level number versus bias voltage for the spectra obtained at different conditions
as noted in the legend. The solid red line is an expected curve calculated by Eq.~\ref{eq1}.
The black dashed line is a linear guide line.}
\label{Fig2}
\end{figure*}
Compared with the experimental results in Fig.~\ref{Fig1}(c), the humps a) and b) are well reproduced.
Especially, hump a) is likely due to the van Hove singularities of the surface-state bands
marked by the black arrow in Fig.~\ref{Figcal}(a). Another hump marked as c) appears
around $-$60~mV in our measurements, but this feature could not be captured by our calculation. On the other hand,
our STS measurements show that its energy level and intensity are not as robust as other features.
These observations indicate that hump c) belongs to a novel surface property.
Based on previous studies~\cite{Takane,Xu}, one can understand the hump c) in terms of couplings of surface electrons with some collective
modes. Here, the collective mode is likely to be phonons on the top layer~\cite{Takane},
which is also observed in other compounds~\cite{WangZY}. Due to the interplay of this collective
mode with the surface states, the LDOS is largely modified.
As a result, the calculated DOS is very different from the STS spectrum around $E_F$.
The correlation of electrons and phonons observed
here provides a new clue for the unconventional mass enhancement in ZrSiS~\cite{Pezzini}.
From Fig.~\ref{Figcal}(b), we also notice that there is a big hump around $E_F$ in our calculation.
It is easy to find that large part of this hump comes from the bulk states when comparing with the
dispersion in Fig.~\ref{Figcal}(a). However, the bulk states are not captured by STS.

By ramping the magnetic field up to 12~T with the $B \parallel c$-axis,
we detected sharp oscillations in the LDOS. As shown in the inset of Fig.~\ref{Fig1}(c),
the oscillations become very obvious after subtracting a smooth background signal
(the background signal is derived by averaging our raw data within $\pm$100 mV with the Savitzky-Golay method).
Typically, these oscillations in th LDOS originate from the Landau quantization
of the electronic states in a magnetic field~\cite{Morgenstern}. In a
2D system, Landau levels are well defined, while in a 3D system they will be
broadened by the $k_z$-component of the electronic states.
Additionally, STM is a surface-sensitive measurement. Therefore, the Landau
levels of the 2D Rashba-split surface states in HfSiS are prone to be detected,
and they contribute to the oscillations in Fig.~\ref{Fig1}(c).
This interpretation is further verified by the following analysis:
The Landau levels show the sharpest peaks near $E_F$ and the
oscillation amplitude fades quickly away from $E_F$. This is related to two effects: 1) the
quasiparticle lifetime $\tau$ of the surface electrons decreases away from $E_F$
($\tau$ $\propto$ 1/($E$-$E_F$)$^2$) \cite{Soumyanarayanan2015}, and 2) with increasing bias voltage,
the tunneling current increases monotonically, which gives rise to a reduced signal-to-noise ratio in STS.
Around $-$60~meV, the aforementioned collective mode further complicates this picture by strongly
smearing the Landau levels, which is similar to the observations in some topological insulators~\cite{Hanaguri,Cheng,Jiang,Okada}.
Again, this phenomenon indicates a surface origin of the Landau levels.
Around $-$80~meV, additional 2-3 Landau levels can be resolved but soon become invisible at lower bias voltage.
Based on the peak width of the Landau levels around $E_F$, we estimate the mean free path $l$ $\approx$ 45 nm
which is comparable with that of the Rashba surface states of pure Sb~\cite{Soumyanarayanan2015} but much shorter than that of
the topologically protected surface states, as e.g. $l$ $\approx$ 80 nm in Sb$_2$Te$_3$~\cite{Jiang}.

Defects in a sample provide a way to study the robustness of the surface states and their quantization.
As shown in Fig.~\ref{Fig2}(a), we measured the STS
spectra along a straight line that crossed three different types of defects (according
to the morphology, we mark these defects as $\bigoplus$, $\square$, and $\maltese$ from the left
to the right side). Figure~\ref{Fig2}(b) presents a contour plot of the d$I$/d$V$-curves that were measured
at 20 equally spaced points along the 20 nm white dotted line in Fig.~\ref{Fig2}(a).
The topography shows that the perturbation induced by $\bigoplus$ extends over several unit cells,
while $\square$, and $\maltese$ are relatively local. Consequently at position $\bigoplus$ the
Landau levels are completely smeared out as evidenced by the uniform color (no oscillation in d$I$/d$V$-intensity)
around curve No.7 in Fig.~\ref{Fig2}(b). On the other hand, Landau quantization can be detected close to the other defects.
Generally, the Rashba parameter is sensitive to the local symmetry and electronic environment,
so the extended wavefunction at position $\bigoplus$ appears to change the value of the Rashba parameter and consequently
modifies the Landau levels. In Fig.~\ref{Fig2}(c),
we present d$I$/d$V$-curves at selected sites after subtracting background signals.
Curves No.1 and No.4 are relatively close to $\bigoplus$ while No.18 and No.19 are around $\maltese$.
At $E_F$, all curves show a maximum (peak), indicating identical phase. Away from $E_F$,
e.g. around $-$26~mV (marked by the dashed line), there is a $\pi$/2 difference in the phase between No.1, No.4 and No.18, No.19:
curves No.1 and No.4 show a minimum (5$\frac{1}{2}$ periods between 0 and $-$26~mV) while No.18 and No.19
show a maximum (6 periods between 0 and $-$26~mV). The latter curves, No.18 and 19, 
are very similar to those measured at clean sites, cf. Fig.~\ref{Fig1}(c)). Taking into account that the peak width of a Landau level is
inversely proportional to its quasiparticle lifetime, the increase in the peak width of curves
No.1 and No.4 indicates notable scattering of the impurity at position $\bigoplus$ to
the surface electrons up to a distance of about 5~nm \cite{Bu}.
We notice that a similar modulation of the phase of the Landau level was detected in the topological insulator
Bi$_2$Te$_3$ around a one-dimensional potential~\cite{Okada2012}.
Figure~\ref{Fig2}(b) also reveals that the collective mode around $-$60~mV is
sensitive to defects at positions $\bigoplus$ and $\square$, while robust against $\maltese$.

For a linear dispersive band, the Landau level ($n$) follows the prediction:
\begin{equation}
E_n(B)=E_C+\nu\sqrt{2e\hbar nB},
\label{eq1}
\end{equation}
where $E_C$ is the band crossing point or the energy level at the high-symmetry point, and $\nu$ is the
Fermi velocity of the surface electrons. The dispersion of the Rashba-split surface states in
HfSiS was found to be quasi-linear away from $E_C$ (both our calculations and ARPES
measurements~\cite{Takane,Chen} indicate $E_C$ $\approx$ $-$400 meV in HfSiS).
Therefore, Eq.~\ref{eq1} can be employed to analyze the Landau levels around $E_F$.
We notice that the Zeeman splitting is neglected due to its small contribution
($\sim$1 meV) compared to the large Rashba-type SOC-induced
splitting. Based on the band structure calculation and ARPES measurements~\cite{Takane,Chen}, one can reasonably
assume that $\nu$ is somewhat isotropic in the momentum space and also in the energy range discussed here.
With the information of $E_C$ as given above, at a fixed magnetic field the free parameters in Eq.~\ref{eq1} are $n$ and $\nu$.
Unfortunately, the 0$^{th}$ Landau level could not be detected in the current study. Nonetheless, $n$ and $\nu$ can be determined by
fitting the experimental data to Eq.~\ref{eq1}.
In Fig.~\ref{Fig2}(d), we plot the peak positions of the Landau levels versus bias voltage.
The data can be nicely fitted to Eq.~\ref{eq1} by using a least-squares method,
which yields $\nu$ = 3.66$\pm$0.05 eV$\cdot${\AA} and $n$ = 33$\pm$1 at $E_F$.
The derived Fermi velocity is in reasonable agreement with the calculated result (4.39 eV$\cdot${\AA}), albeit somewhat smaller.
The reduction of $\nu$ may be attributed to the influence of the collective mode, which slightly bends the surface-state band away
from a perfect linear dispersion~\cite{Takane} and consequently induces some error when using Eq.~\ref{eq1}.
This observation well resembles the mass enhancement observed by measuring Landau level spectroscopy in Pb$_{1-x}$Sn$_x$Se~\cite{Zeljkovic},
in which the electron-phonon coupling plays a role. The dashed black line
in Fig.~\ref{Fig2}(d) shows clear deviations of the Landau levels from a linear behavior,
which is expected for a normal parabolic dispersive band.

\begin{figure}[tb]\centering
\includegraphics[width=7cm]{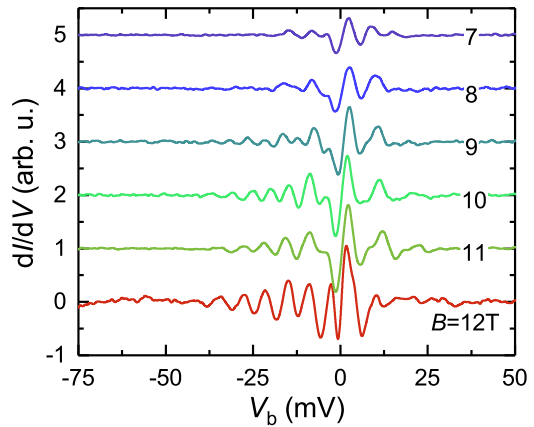}
\caption{ Magnetic field dependent d$I$/d$V$-curves with a smooth
background subtracted. These curves are measured at clean surface from 7 T to 12 T at $T$ = 0.35~K.
Curves are offset vertically for clarity.}
\label{Fig3}
\end{figure}

Figure~\ref{Fig3} presents the Landau quantization in STS spectra for various magnetic fields,
with a background signal subtracted. Landau levels are visible from 7 T.
With increasing magnetic field $B$, both the height and the width of the Landau level peaks
are expected to increse by Landau quantization, which is nicely manifested in Fig.~\ref{Fig3}.
\begin{figure*}[tb]\centering
\includegraphics[width=15cm]{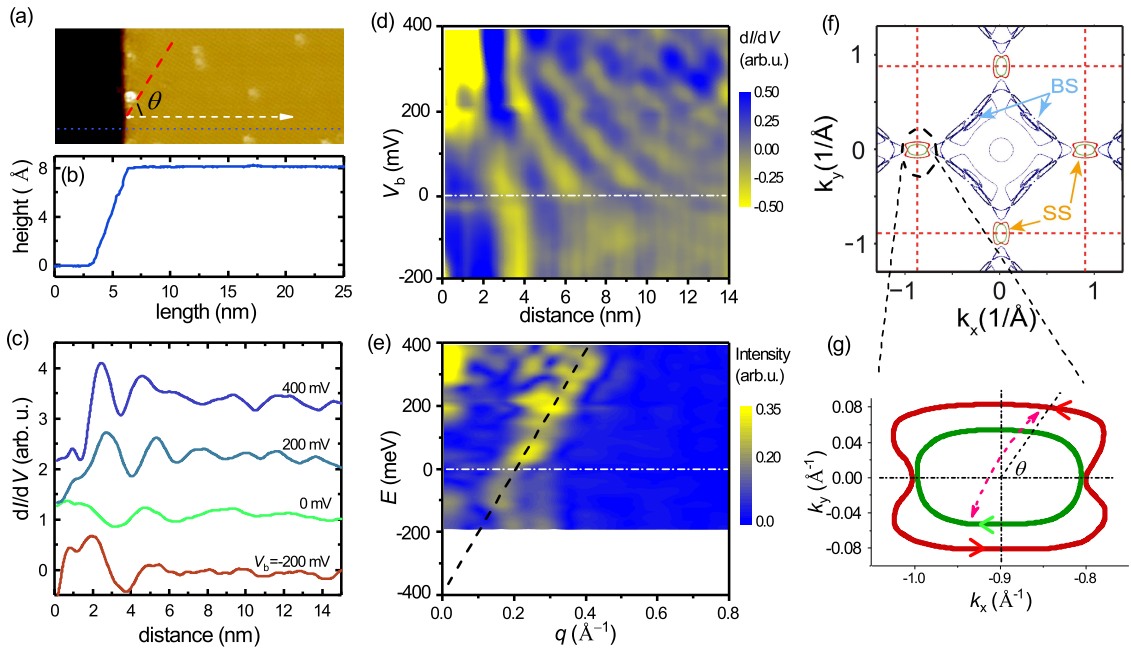}
\caption{(a) 25$\times$10 nm$^2$ topography of HfSiS with a sharp step edge. $\theta$ $\approx$ 55$^\circ$ is the angle between the
crystal axis (red dash line) and the normal of the step edge (white arrow). (b) A line
scan (the blue dot line in (a)) shows the height of the step edge is of one unit-cell (8 {\AA}).
(c) d$I$/d$V$-curves measured at selected bias voltages along the white dashed line in (a).
Data was taken at 0.35~K and 0~T with
$V_b$ = 0.2~V and $I_{set}$ = 0.4~nA. Curves are equally shifted for clarity.
(d) A contour plot of the d$I$/d$V$-map measured along the white dashed line
at various bias voltages. (e) A Fourier transform of the d$I$/d$V$-map in (d). The
black dashed line is a linear extrapolation of the frequencies of the standing waves.
(f) Reproduced Fermi surface of HfSiS reported in our previous calculation~\cite{Chen}.
Bulk states (BS) and Rashba surface states (SS) are marked in the plot.
(g) A sketched Fermi surface of the Rashba-split
surface bands around one of the four $\overline{X}$-points, which is based on the calculation in (f).
The pink arrow indicates a possible scattering vector on the Fermi surface with 55$^\circ$ off $k_x$-direction.
Red and green color mark the opposite spin orientations of the outer and inner surface-state bands.}
\label{Fig4}
\end{figure*}

In addition to real-space analysis, STM can also be employed to obtain momentum-space
information via the modulation of the LDOS around a point defect or a step edge. Along an atomically sharp edge,
the surface states can be scattered backward by the potential barrier/well, which results in an
interference pattern containing wave-vector information of the surface electrons. This picture is based on the
so-called Friedel oscillation~\cite{Friedel,Hasegawa,Sessi2015}. In particular, surface states without a $k_z$
component of momentum are prone to be backscattered and form the standing-wave patterns \cite{Crommie,Sprunger}.
For simplicity, the so-obtained patterns themselves are often referred
to as Friedel oscillations~\cite{Hasegawa,Sessi2015,Sprunger}.
Figure~\ref{Fig4} shows the standing waves observed near a step edge on HfSiS.

The height of the step edge is around 8~{\AA}, which is one
unit cell. By measuring the STS along the white dashed line in
Fig.~\ref{Fig4}(a), we get a d$I$/d$V$-map versus bias conductance around the step edge.
Figure~\ref{Fig4}(c) shows the d$I$/d$V$ intensity versus distance at several selected energies.
Oscillations of the LDOS can be resolved directly from the raw data,
especially at high bias voltage. To illustrate the evolution of the oscillation frequency versus
bias voltage, in Fig.~\ref{Fig4}(d), we provide a contour plot of the d$I$/d$V$-map.
It is obvious that the wave length decreases (therefore the oscillation frequency increases)
with increasing bias voltage, which is further manifested in the Fourier transform of
the d$I$/d$V$-map, i.e. Fig.~\ref{Fig4}(e). In this plot a broad dispersion
line is resolved, which represents the length between two wave vectors
in the momentum space in dependence on energy, known as $q$-space.
With increasing energy, the frequency increases linearly, being in agreement with the band structure
of the Rashba-split surface states in Fig.~\ref{Figcal}(a).
In this respect, Fig.~\ref{Fig4}(g) shows one of the possible wave vector between two
$k$-points with the same spin orientation, the direction
of which is the same as the step-edge (55$^\circ$ off $k_x$-direction).
However, such standing waves could not be detected below $-$100~meV, which is likely
due to the perturbation of the collective mode. However, a linear extrapolation of the
oscillation frequency indicates $E_C$ is around $-$400 meV [Fig.~\ref{Fig4}(e)], which is consistent with our
calculation as well as the ARPES measurements~\cite{Chen,Takane} of the Rashba-split
surface states. Additionally, based on the slope of
the linear dispersion in Fig.~\ref{Fig4}(e), we estimate that the Fermi velocity of the Rashba-split surface states is
around 4.0(5)~eV {\AA}, which is close to the results obtained from the Landau quantization.

In summary, we investigate the Rashba-type surface states of Dirac nodal line semimetal HfSiS by utilizing STM down to 0.35~K
and up to 12~T. Landau quantization of the surface states can be observed above 7~T, which yields a Fermi
velocity of $\sim$3.66~eV$\cdot${\AA} and a mean-free-path of $\sim$45~nm. On the other hand, the dispersion of the Rashba-split
surface states is also studied by measuring the standing waves along a step edge,
which also indicates a strong spin polarization of these surface states.
We further show that different defects have different effects on the surface states. All the experimental results are
quite consistent with our band structure calculation. These detailed surface properties of HfSiS will
provide insight into our understanding of this new type of semimetal.

\acknowledgments
We acknowledge valuable discussion with Y.P. Jiang, D. Kasinathan, and G.B. Zhu.
We specially thank S. R\"o{\ss}ler for the valuable comments on the manuscript.
Financial support from the Deutsche Forschungsgemeinschaft (DFG)
within the Schwerpunktprogramm SPP1666 is gratefully acknowledged.
L.J. acknowledges support by the Alexander-von-Humboldt foundation.

\end{document}